\documentclass[12pt,onecolumn] {article} 
\usepackage{amsmath}
\usepackage{amssymb}
\usepackage{latexsym}
\usepackage{graphicx}

\begin{document}

\begin{center}
\baselineskip=24pt

{\Large \bf Measurements of neutrons produced by high-energy muons
at the Boulby Underground Laboratory}

\baselineskip=18pt

\vspace{0.3cm}
H.~M.~Ara\'ujo$^{a,b}$,
J.~Blockley$^{c}$,
C.~Bungau$^{b,a}$,
M.~J.~Carson$^{c}$,
H.~Chagani$^{c}$,
E.~Daw$^{c}$,
B.~Edwards$^{a,b}$,
C.~Ghag$^{d}$,
E.~V.~Korolkova$^{c}$,
V.~A.~Kudryavtsev$^{c}$ \footnote{Corresponding author; address: 
Department of Physics and Astronomy, University of Sheffield,
Sheffield S3 7RH, UK, e-mail: v.kudryavtsev@sheffield.ac.uk},
P.~K.~Lightfoot$^{c}$,
A.~Lindote$^{e}$,
I.~Liubarsky$^{b,a}$,
R. L\"{u}scher$^{b}$,
P.~Majewski$^{c}$,
K.~Mavrokoridis$^{c}$,
J.~E.~McMillan$^{c}$,
A.~St.~J.~Murphy$^{d}$,
S.~M.~Paling$^{c}$,
J.~Pinto da Cunha$^{e}$,
R.~M.~Preece$^{b}$,
M.~Robinson$^{c}$,
N.~J.~T.~Smith$^{b}$,
P.~F.~Smith$^{b}$,
N.~J.~C.~Spooner$^{c}$,
T.~J.~Sumner$^{a}$,
R.~J.~Walker$^{a}$,
H.~Wang$^{f}$ and
J.~White$^{g}$

$^{a}$ {\it Blackett Laboratory, Imperial College London, UK}\\
$^{b}$ {\it Particle Physics Department, STFC Rutherford Appleton 
Laboratory, UK}\\
$^{c}$ {\it Department of Physics \& Astronomy, University of Sheffield, UK}\\
$^{d}$ {\it School of Physics, University of Edinburgh, UK}\\
$^{e}$ {\it LIP--Coimbra \& Department of Physics of the 
University of Coimbra, Portugal}\\
$^{f}$ {\it Department of Physics \& Astronomy, University of 
California, Los Angeles, USA}\\
$^{g}$ {\it Department of Physics, Texas A\&M University, USA}\\

\vspace{0.3cm}
\begin{abstract}

We present the first measurements of the muon-induced neutron flux at the Boulby
Underground Laboratory. The experiment was carried out with an 0.73~tonne
liquid scintillator that also served as an anticoincidence system for the ZEPLIN-II
direct dark matter search. The experimental method exploited the delayed coincidences 
between high-energy muon signals and gamma-rays from radiative neutron capture on
hydrogen or other elements.
The muon-induced neutron rate, defined
as the average number of detected neutrons per detected muon, was
measured as $0.079 \pm 0.003$ (stat.) neutrons/muon using
neutron-capture signals above 0.55~MeV in a time
window of 40-190~$\mu$s after the muon trigger. 
Accurate Monte Carlo simulations of the
neutron production, transport and detection in a precisely modeled laboratory and
experimental setup using the GEANT4 toolkit gave a result 1.8 times 
higher than the measured value. The 
difference greatly exceeds all statistical and systematic uncertainties.
As the vast majority of neutrons detected in the current setup were produced in lead
we evaluated from our measurements the neutron yield in lead as
$(1.31 \pm 0.06) \times 10^{-3}$ neutrons/muon/(g/cm$^2$) 
for a mean muon energy of about 260~GeV.

\end{abstract}

\end{center}

\pagebreak

\section {Introduction}
\label{intro}

Background from muon-induced neutrons is one of the most important limitations
to detector sensitivity for rare event searches and is largely responsible
for the existence of the field of physics called `Underground Physics' and of
deep underground laboratories across the world. 

In WIMP dark matter detectors nuclear recoils of keV energies originating in 
neutron elastic scattering mimic WIMP-nucleus interactions. 
For double-beta decay experiments high-energy neutrons (about a few MeV
and above) produce
background gamma-rays via inelastic scattering while thermal neutrons
contribute to the gamma-ray background via neutron capture accompanied
by gamma-ray emission. Neutrons at MeV energies and above also mimic
neutrino detection in scintillators via inverse beta decay posing a severe
threat to low-energy neutrino experiments (reactor and geo neutrinos). 
Neutrons at sub-GeV and GeV
energies, although rare, constitute the background for proton decay
and atmospheric neutrino experiments.

Neutrons from radioactivity originate in spontaneous fission 
(of $^{238}$U mainly) and ($\alpha$,n) reactions on low and intermediate
Z isotopes (Z$\lesssim$30). Their energies are limited to about 10~MeV. 
Neutrons from cosmic-ray muons have spectra extending to GeV energies.
Although the flux of muon-induced neutrons deep underground
is far below that from radioactivity, they can be responsible for a 
significant background component and limit sensitivity of detectors 
to rare events due to the following reasons. Firstly, neutrons from radioactivity
in rock can be well shielded by hydrogen-rich material, while using ultra-pure
materials in detector construction reduces the background from all other 
components. For muons, however, any material in shielding or detector 
(except hydrogen) is a 
target for neutron production. Secondly, as muon-induced neutrons have higher
energies, it is much more difficult to moderate and/or absorb them. They 
can travel far from the muon track or their point of origin reaching detectors
from large distances and reducing the efficiency of an anticoincidence
system. Any high-A material is also a good target for secondary neutron 
production from primary neutrons. All this makes the measurements 
and calculation of muon-induced
neutron rate an important task for designing and constructing sensitive
detectors for rare event searches.

Neutrons are produced by muons via 4 main processes:
i) negative muon capture (relevant only to low-energy, stopping
muons, or for shallow depths less than 100 m~w.~e.);
ii) direct muon-induced spallation of a nucleus;
iii) photoproduction of neutrons or photon-induced spallation 
(mainly in electromagnetic cascades initiated by muons);
iv) hadroproduction of neutrons (mainly in hadronic or nuclear
cascades originated by muons).
The relative contribution of different processes in different models have been
investigated in Refs.~\cite{vak03,ha05}.

There were many attempts
to measure the neutron fluxes produced by high-energy muons in the
laboratories at surface (using accelerators) and underground 
(see, for example, 
\cite{bezrukov,asd,lsd,lvd,paloverde,chen,menghetti04,na55,
gorshkov74,bergamasco73}) and more experiments are planned \cite{akerib}. 
Reliable simulations of the expected effects have become possible
only recently with the appearance of powerful Monte Carlo codes based on 
advanced theoretical models, such as FLUKA \cite{fluka} and GEANT4 
\cite{geant4}. Several measurements of neutron
yield in liquid scintillator at different depth 
\cite{bezrukov,asd,lsd,lvd,paloverde,chen}
are in agreement with
FLUKA \cite{fluka} and GEANT4 \cite{geant4} within a factor of two 
or better as discussed in 
Refs. \cite{wang01,vak03,ha05}. Three of them 
carried out at depths more than 50 m~w.~e. \cite{bezrukov,asd,lsd}
show higher neutron
yield than FLUKA and GEANT4 predictions.
The LVD experiment \cite{lvd} reported smaller neutron yield in scintillator 
than predicted
by either FLUKA or GEANT4. Neutron yield in lead has been reported in
Ref. \cite{bergamasco73} as 0.016 n/$\mu$/(g/cm$^2$) for a mean muon
energy of about 310 GeV. This value is two (four) times higher than simulations
carried out with the FLUKA (GEANT4) code \cite{ha05}. Significant
excess of neutrons in lead over model predictions was observed also in Ref.
\cite{na55}. 

The evaluation of neutron yields
from the experimental data, however, is not straightforward
and requires, in its turn, detailed Monte Carlo simulation of the setup
and all physical processes involved.
This is difficult to do now for early experiments since
not all details of the setups are known. At the time when these experiments
were carried out, the theoretical models and computer codes were
not developed to the extent that would allow accurate simulations of
the expected effects. So any comparison between old data and simulations
should be taken with caution. Hence there is an urgent need for 
new experimental data on neutron yields in different materials, as well as
neutron energy spectra, lateral distributions etc. supported by accurate
Monte Carlo simulations with widespread multi-purpose codes or toolkits.

In this paper we describe the measurement of the total neutron yield
from cosmic-ray muons carried out at the Boulby Underground Laboratory
(Boulby mine, North Yorkshire, UK).
This is the first experiment with a large mass of lead as a target for which
direct measurements and detailed Monte Carlo are compared directly 
enabling the test of the models. Our results are relevant to many
sensitive underground experiments for rare event searches that use
or will use lead as a shielding against gamma-rays from rock and hence
expect a large background from neutrons produced by muons and their
secondaries in lead.

Due to the relatively simple setup, the systematic uncertainties from 
geometry, trigger effects etc. are reduced to the minimum allowing,
for the first time, accurate calculations of the expected neutron rate 
using GEANT4 version 8.2.

In Section \ref{detector} we describe the detector and data
acquisition used in the
measurements of the muon-induced neutron flux. Section
\ref{results} shows our results. Monte Carlo simulations
of the detector setup and physical processes are presented
in Section \ref{sim}. We discuss and compare our results
to simulations in Section \ref{discussion}. The conclusions
are given in Section \ref{conclusions}.

\section {Detector, data acquisition and data analysis}
\label{detector}

The measurements were carried out at the Boulby Underground Laboratory
at a depth of about 1070~m or 2850~m~w.~e. The liquid scintillator
of an active veto system working also
in anticoincidence with the WIMP dark matter detector ZEPLIN-II 
\cite{z2} was used for muon and neutron detection. 

\begin{figure}[htb]
   \includegraphics[width=15.cm]{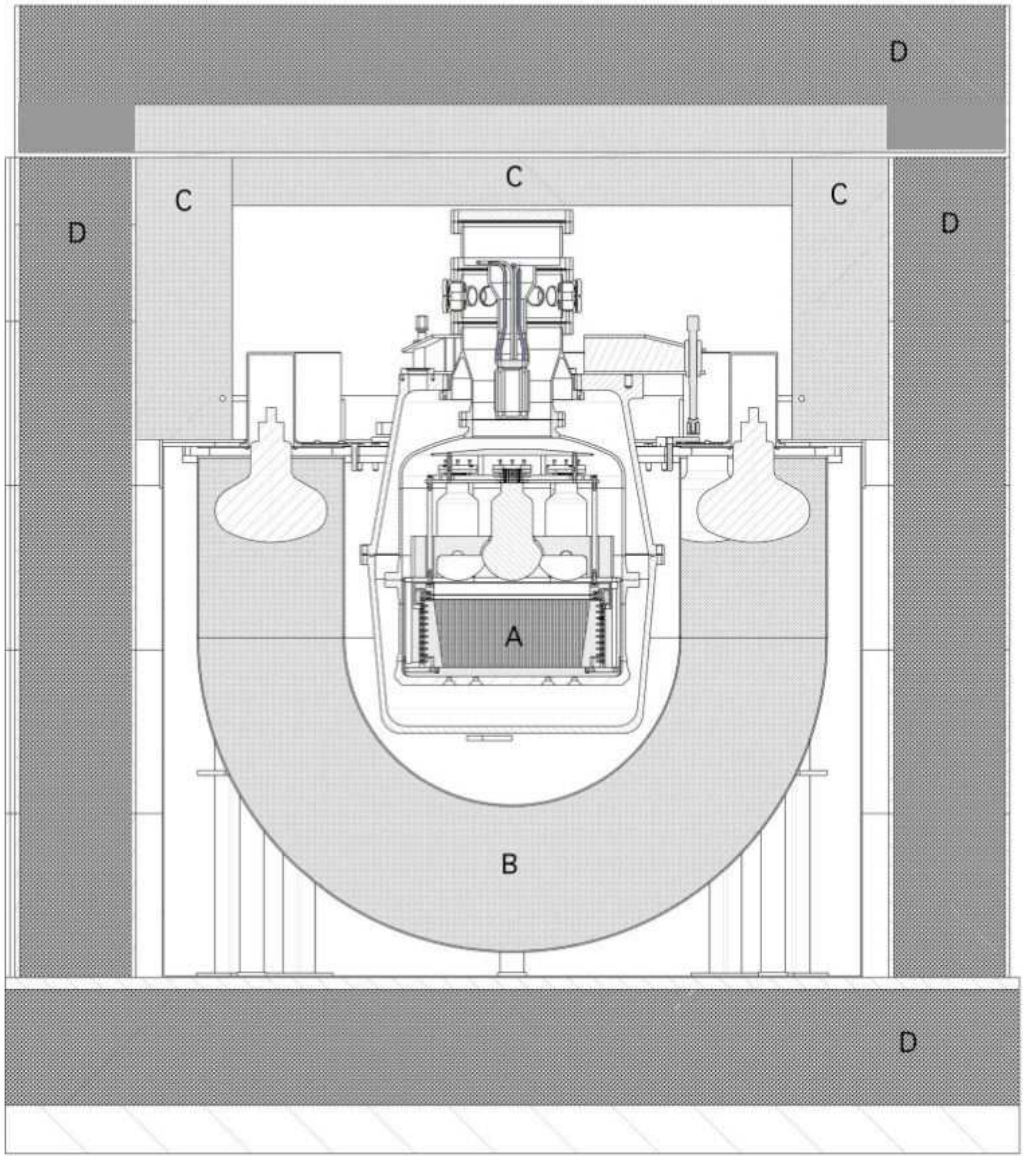}
    \caption{Schematic of the veto system used in neutron measurements, 
    together with shielding and the ZEPLIN-II detector:
    A -- ZEPLIN-II detector, B -- liquid scintillator used in the present
    measurements, C -- Gd-loaded wax, D -- lead (reproduced from \cite{z2}).}
  \label{fig-veto}
\end{figure}

The veto detector (Figure~\ref{fig-veto}) is a hollow structure 
surrounding the 
ZEPLIN-II experiment on five sides. The lower part of the detector is
hemispherical in shape with an inner radius of 0.35~m and an outer radius 
of 0.65~m giving a distance between walls of 0.3~m. 
The upper part is cylindrical in shape with the same inner and outer radii 
and a height of 0.36~m. The vessel is filled with liquid scintillator based on 
mineral oil containing approximately 25\% of phenyl-o-xylylethane (produced
by Elgin).
This was chosen primarily for its high flash point due to safety
constraints on materials used at Boulby. The scintillator has
a density of 0.89~g/cm$^3$, average atomic number of 4.75 and
average atomic weight of 8.33. The wavelength of maximum emission is
425~nm and the attenuation length exceeds 2~m. 
The light output, as specified by the manufacturer, is 57\% of that for
anthracene. Our calibration carried out using a $^{60}$Co gamma-ray
source and single photoelectron pulses, revealed a light
yield of about 30 photoelectrons per MeV.
The volume of the detector is 
0.82~m$^3$ giving a total mass of liquid scintillator of 0.73~tonnes.
The scintillator is viewed from above by ten 20~cm ETL hemispherical
photomultiplier tubes (PMTs).
The inner surfaces of the detector are covered in
aluminium with a coefficient of reflection greater
than 0.9 to maximise light collection.
More details about the veto detector and its performance relevant
to dark matter searches with ZEPLIN-II can be found in 
\cite{z2}.

The veto detector (together with the ZEPLIN-II liquid xenon detector)
is surrounded by a `castle' made of lead and designed to shield
the ZEPLIN-II dark matter detector from gamma-rays from rock.
The thickness of the lead shielding ranges from 15~cm
on top to 22.5~cm below the veto and on four sides (Figure~\ref{fig-veto}).
The total weight of lead is about 50~metric tonnes making it an
excellent target for neutron production.

Another important feature of the setup is the presence of pure and
Gd-impregnated wax and polypropylene (about 0.2\% of Gd
by weight on average) on top of the veto detector 
under the castle roof. The inner surface of the veto vessel (close to the ZEPLIN-II
detector) was also covered with a paint mixed with Gd salt.
The purpose of this was to shield the ZEPLIN-II target from neutrons
from rock. As we see later, Gd did not
help in detecting neutron capture events in the time window used in our measurements. 
The conclusion may be different
for the ZEPLIN-II dark matter detector where the main neutron 
background is assumed to be originated just around the detector
(PMTs, stainless steel vessel etc.).

Part of the data run was carried out without the top ('roof') lead and wax sections, reducing
the total lead target mass by about 12\%. This did not have a large impact
on neutron production: the neutron event rate was slightly reduced
for the time periods without the roof section. The relatively small reduction
can be explained by the fact that wax on top of the detector, that could 
efficiently absorb neutrons,
was also removed in those runs together with lead.

All materials of and around the veto detector and their exact locations
within the underground laboratory
were put into the simulation code based on GEANT4 toolkit
\cite{geant4} as will be described later.

The principle of neutron detection was based on the delayed
coincidences between the first, high-energy, pulse from a 
muon or muon-induced cascade and the delayed second, low-energy,
pulse from neutron capture gamma(s). The electronics and data acquisition
system were designed for this purpose.

Signals from the ten PMTs were passed through discriminators and 
a coincident unit. The unit generated a logic pulse if the amplitudes
of the analogue pulses from three or more PMTs exceeded
an average amplitude of one photoelectron within 150~ns.
The logic pulses were used in the off-line analysis to identify
`true' pulses and reject noise and other background pulses
caused by radioactivity close to a single PMT (for instance on the
PMT window) without coincidences with any other PMT.

Analogue signals from all ten PMTs were summed together by an adder and 
the sum signal was fed into a waveform digitiser sampling
at a rate of 500~MHz.
The sum pulse was digitised at two different amplitude ranges: 0.5~V and 5~V.
The first range was used primarily for low-energy pulses from neutron capture
gamma-rays, while the large range was important for identifying
muon events. The adder had a range of about 2~V. All pulses with amplitude
exceeding this limit were truncated. The gamma-ray spectrum from natural radioactivity
extends to a few MeV corresponding to less than 1~V (the energy calibration 
procedure is described in Section \ref{results}). Taking into account
the non-uniform light collection, the gamma-ray spectrum extends to almost
2~V in amplitude. The minimum thickness of the scintillator in the vertical direction
is 30~cm or 26.7~g/cm$^2$. A muon crossing this thickness of scintillator
deposits on average about 50~MeV which is far above the saturation
level of the adder. So practically all muon pulses were saturated (truncated)
with the exception of some fraction of events with muons cutting edges of the
detector or events caused by low-energy secondaries. 
This feature was used in the selection procedure for muon events.

In the `data' run the trigger to the DAQ 
was provided by large analogue
pulses exceeding about 1.2~V. This corresponded approximately to
10~MeV energy deposition in the detector. The rate of recorded events
in the data run was a few events/hour consisting of muons and a tail of
gamma events seen at higher energies due to their 
location close to PMTs.
The 5~$\mu$s and 195~$\mu$s time periods before and after the 
trigger pulse, respectively, were digitised allowing for the delayed pulse 
from neutron capture to be recorded.

The logic pulse, generated by the coincidence unit if at least three
PMT hits were above the threshold equivalent to an average amplitude
of a single photoelectron, was also recorded on the same waveform
100~ns after the start of the 
analogue pulse. Only analogue signals accompanied by logic pulses were
considered in the off-line data analysis.

Event waveforms were parameterised using the data reduction code 
similar to those written for the neutron measurements from radioactivity in 
rock described in Ref. \cite{nuboulby} and for the ZEPLIN-II experiment
\cite{z2-daq}. For each pulse on the waveform, the area (proportional
to the charge and, hence, deposited energy), amplitude, width,
mean time, full width at half maximum (FWHM) and arrival time were
recorded. In addition, the charge 
in the first 20~ns from the beginning of the pulse and the time during
which the amplitude of the pulse exceeded a pre-defined threshold of
1.7~V were calculated. The last parameter was particularly important
for identifying muon events since an accurate measurement of the
muon energy deposition was impossible due to the adder saturation.
(Even without this effect it would be impossible for PMTs and DAQ 
to cover with equal precision the MeV range relevant for gamma-rays
from neutron capture and tens of MeV or higher energy depositions 
from muons and cascades.)

Only pulses that exceeded 0.02~V in amplitude 
(equivalent to approximately 200~keV energy deposition) and were
accompanied by a logic pulse, were selected for further analysis.

\begin{figure}[htb]
   \includegraphics[width=15.cm]{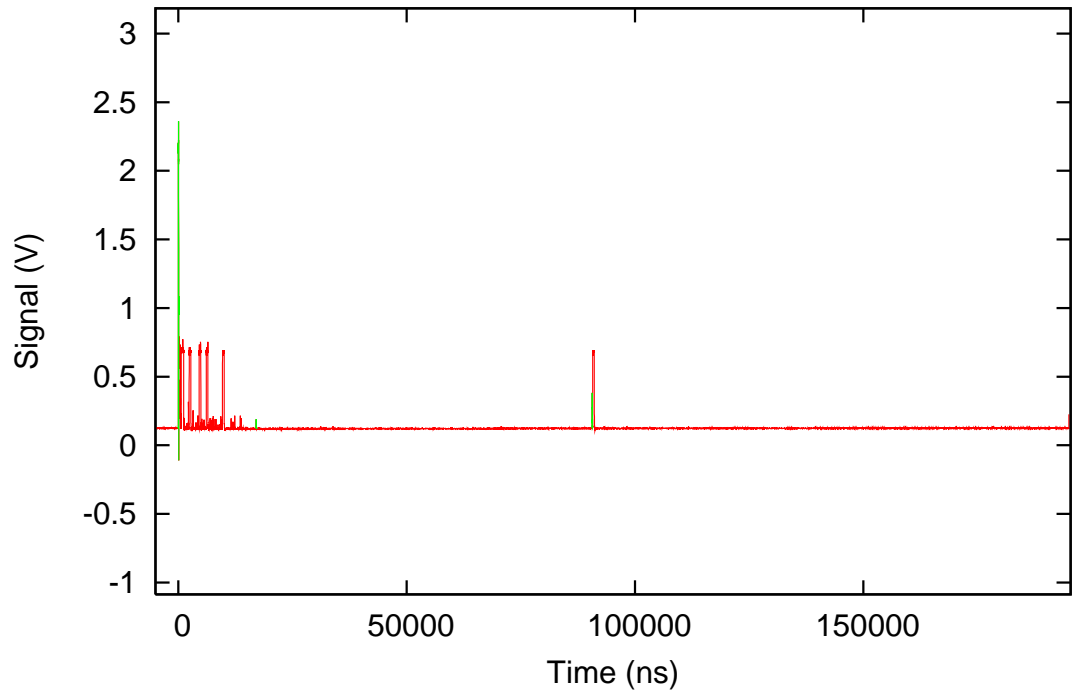}
    \caption{Typical muon event with a neutron-like pulse (from neutron capture) 
    that occurs at about 90~$\mu$s after the muon trigger.
    Pulses with about 0.6~V amplitude are logic pulses as explained
    in the text.}
  \label{fig-pulse}
\end{figure}

\begin{figure}[htb]
   \includegraphics[width=15.cm]{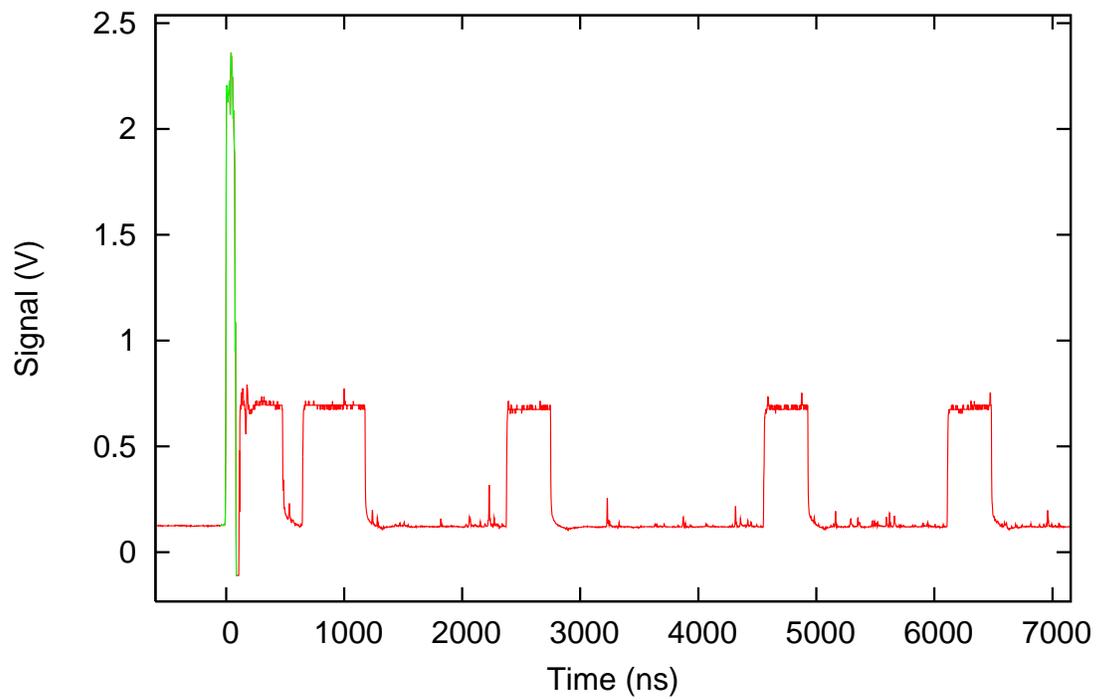}
    \caption{Magnified view of the trigger region for the same event as in 
    Figure \ref{fig-pulse}.
    Pulses with about 0.6~V amplitude are logic pulses. Muon events are
    characterised by a large number of afterpulses accompanied by logic pulses
    preventing reliable neutron detection 20~$\mu$s after the trigger.}
  \label{fig-muon}
\end{figure}

Figure~\ref{fig-pulse} shows a typical waveform for a muon-induced
neutron event. Figure~\ref{fig-muon} displays the time period of about
8~$\mu$s
around the trigger. The muon pulse occurs at a trigger time 
$0\pm 100$~ns. It is characterised by a nearly flat top part
(effect of adder saturation resulting in pulse truncation), large width
(both mean time and FWHM) and the presence of many afterpulses
during the first 15-20~$\mu$s after it. Logic pulses are those with a 
width of about 0.4~$\mu$s and an amplitude of 0.6~V. As the dynamic
range of the PMTs was chosen for the best sensitivity to MeV
energy depositions, higher energy muon pulses saturated PMTs
resulting in a large number of afterpulses. Some afterpulses from 
different PMTs coincided in time producing logic 
pulses. The presence of this `forest' of afterpulses made impossible
any analysis of neutron captures that could occur within the first
15~$\mu$s after a muon. Only delayed pulses in the time period
of 15-195~$\mu$s after the trigger were analysed by the data
reduction code. This selection suppresses the detection of gammas from
neutron capture on Gd. The two isotopes of Gd 
have very high neutron capture cross-section and, hence, the small
mean time delay for neutron capture, about 15~$\mu$s for 0.2\% of Gd,
while the analysis of secondary pulses from neutron capture started
only 15~$\mu$s after the muon signal.
A typical pulse from neutron capture gamma-ray(s)
is shown in Figure~\ref{fig-pulse} at a time of about 90~$\mu$s.
Due to the large time window and small amplitude of the analogue pulse,
only the logic pulse can be seen.

From all pulses found on the waveform in the time window
15-195~$\mu$s after the muon trigger, only 20 with highest charge were
recorded. This had some effect on the number of detected neutrons since simulations
showed that multiplicities larger than 20 could also be observed. 
Hence the maximum detected multiplicity is 20 for 15-195~$\mu$s
time window. This effect will be studied later.

\section {Experimental results}
\label{results}

\subsection{Calibrations}

\begin{figure}[htb]
   \includegraphics[width=15.cm]{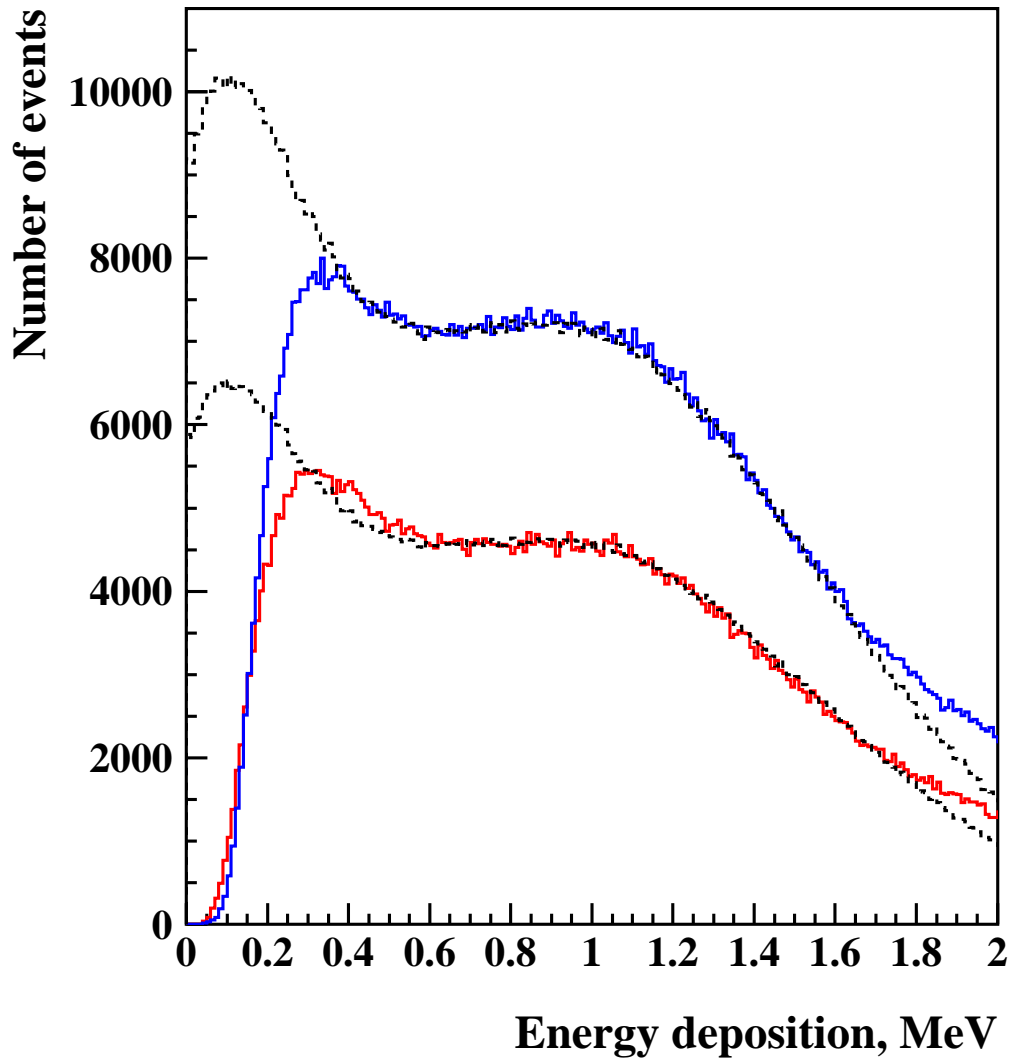}
    \caption{Energy spectra of gamma-events from $^{60}$Co energy calibration runs:
    upper solid histogram -- first calibration in August 2006 before the beginning
    of the data run; lower solid histogram -- second calibration in March 2007,
    close to the end of the data run. Dashed histograms show normalised 
    simulated spectra.}
  \label{fig-encal}
\end{figure}

Two types of detector calibration were carried out: energy calibration with
a gamma-ray source and calibration with a neutron source. 
Energy calibrations with $^{60}$Co gamma-ray source were performed at the
beginning of the long data run and at the end of the experiment. They allowed
us to determine the energy scale for events, namely, the conversion from
the area of a pulse to the energy deposited in the detector.
In the energy calibration run only 1~$\mu$s of the waveform was recorded
since only one pulse without any delayed coincidences was expected
from gamma-rays. The amplitude threshold for the DAQ was reduced to
about 0.4~V allowing triggering on the logic pulse and detecting analogue
pulses with smaller amplitudes. 
Figure~\ref{fig-encal} shows the energy spectra of events from
$^{60}$Co gamma-ray source collected at the beginning and 
at the end of the data run.
No gamma-ray line is seen in the spectra due to poor energy resolution 
(large non-uniformity of
the light collection) of the detector but the shoulders correspond 
to the combination of the two Compton edges and two 
full absorption peaks of the $^{60}$Co gamma-ray lines at 1.173~MeV
and 1.333~MeV. 

The dashed histograms show the normalised
simulated spectra. The simulations were carried out using the GEANT4
toolkit \cite{geant4} taking into account the geometry of the veto and surroundings
and the position of the $^{60}$Co source. 
The energy depositions of photons in scintillator
(through secondary electrons) were recorded. The light collection was not 
simulated but was taken into account by applying a Gaussian 
smearing to the energy deposition in each event. The standard
deviation (or $\sigma$) for the smearing was described by the equation:
$\sigma / E = \sqrt{\alpha + \beta / E}$ suggested by Birks \cite{birks} and
used also in the description of the energy calibration results from
a small scintillator cell for neutron background measurements at
Boulby \cite{nuboulby}.
The parameters $\alpha$ and $\beta$, as 
well as the conversion factor (for the data)
from the measured pulse area to the 
energy scale, were determined from a comparison between the measured 
spectra and simulations.

The two dashed histograms of simulated events differ only by the normalisation
factor that determines the vertical scale. They use the same values for the 
parameters $\alpha$ and $\beta$. The two data sets (solid histograms) are
plotted using slightly different pulse area -- to -- energy conversion factors, namely:
0.36 MeV/(V$\times$ns) for the first run and 0.34 MeV/(V$\times$ns) for the second 
run showing good stability of the detector over several months of running.
The agreement between simulation and both data sets in the energy range of interest
(0.5-1.5 MeV) demonstrates the
reliability of the energy calibration and of the simulation model. The difference
of 6\% in the pulse area -- to -- energy conversion factors between the two data runs
determines the uncertainty in the energy scale.
The difference between data and simulations at low energies is due to the
energy threshold (not simulated). The difference at high energies is probably
due to the strong enhancement of the signal if the energy deposition occurs
close to the PMTs and/or to a possibility of detecting energy
depositions of two photons from the
same decay (not taken into account in simulations).
As will be shown later, more reliable energy calibrations, though with larger 
uncertainty in the energy scale can be obtained directly from the data using
the 2.22~MeV gamma-rays from neutron capture on hydrogen.

Neutron calibration with an Am-Be source of 0.1~GBq $\alpha$-activity
has been carried out before the beginning of the 
data run. 
Ideally, the calibration of the experiment with
a neutron source would allow the evaluation of detector efficiency
to neutrons. The outcomes of calibration carried out in this 
experiment, however, were limited due to the reasons outlined
below. Firstly, neutrons from the Am-Be source have on average 
lower energies than those produced by cosmic-ray muons.
Secondly, neutrons are produced by muons everywhere, while
the source position was fixed (on top of the veto detector,
just below the castle roof). The neutron capture may thus occur on
different materials during the neutron calibration and data runs.
Thirdly, the trigger in the neutron calibration run was different from that
in the data run. In the data run the high-energy 
muon energy deposition was the natural trigger, 
whereas in the neutron calibration run,
in the absence of muons, the low-energy trigger from neutron-induced
proton recoils was used.
Finally, the neutron event rate during the calibration run was a few
tens of events per second while the maximum rate handled by the DAQ
(with almost 100\% dead time) was about 20~Hz (in any run
that recorded 200~$\mu$s waveform for each event).
Hence, the main aims of the neutron calibration were: i) to demonstrate
that the experiment is sensitive to neutrons; ii) to show that the
Monte Carlo models are accurate enough in describing neutron
transport and detection by comparing 
measured and simulated time delay
distributions of neutron-capture signals.

We point out that similar restrictions to calibration accuracy apply to other experiments that 
measured muon-induced neutron flux and calibrated their detectors using
neutron sources.

\begin{figure}[htb]
   \includegraphics[width=15.cm]{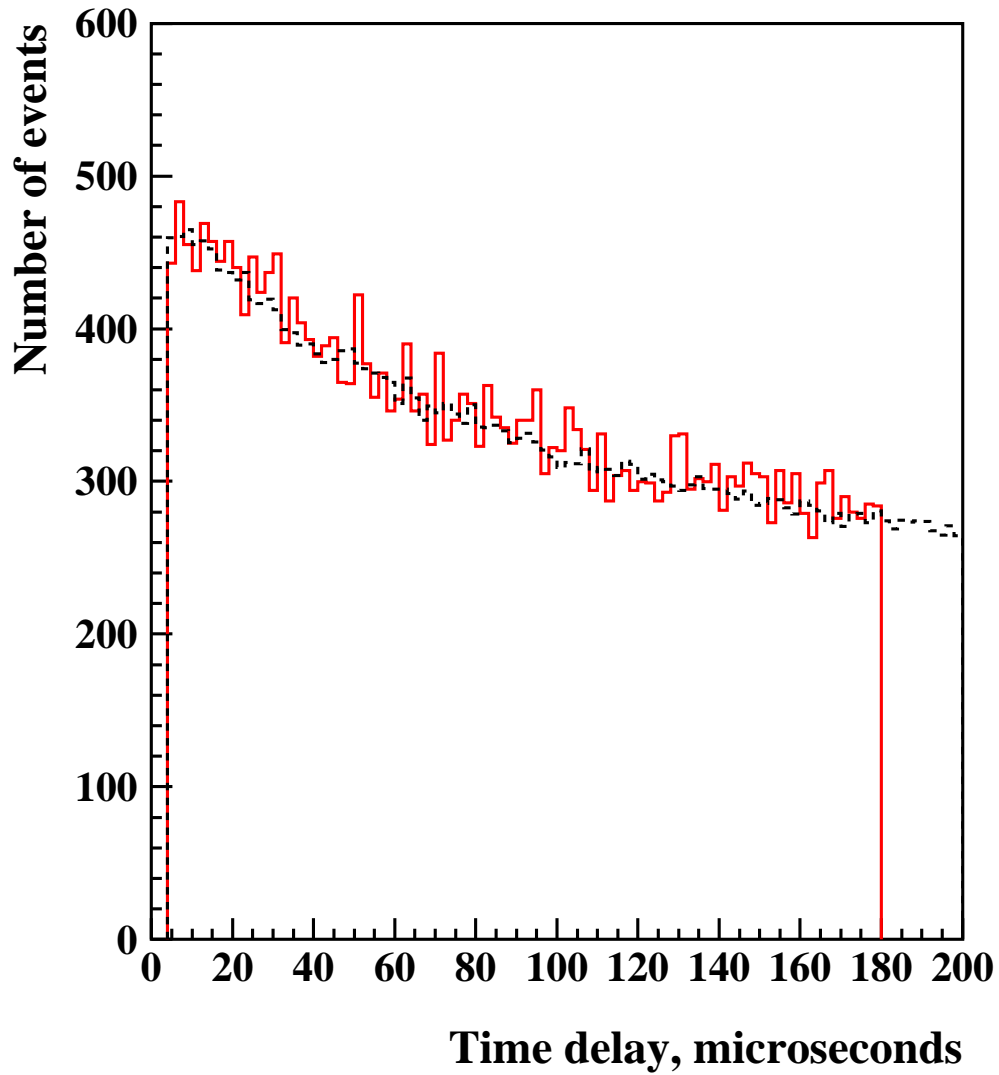}
    \caption{Time delay distribution of the secondary pulses in a neutron
    calibration run with Am-Be source. The data (solid histogram) are plotted together
    with the simulated distribution (dashed histogram, see text for details).}
  \label{fig-ncal}
\end{figure}

Figure~\ref{fig-ncal} shows measured and simulated time delay 
distributions
between the pulses in the events relative to the trigger pulse
in the neutron calibration run.
Simulation of this run was carried out using GEANT4
taking into account the geometry of the setup, position of the source,
neutron interactions and capture. The initial energy spectrum
of neutrons from Am-Be source was calculated using the SOURCES4
\cite{sources4} code. Two free parameters were used to tune the
simulated distribution to match the data: (i) the normalisation constant
or the total number of neutrons; (ii) the flat `background' component
that was seen in the data but was not simulated. 
The natural background during the calibration run was small and
was not responsible for the flat component of the time delay
distribution. Instead, the flat component was due to random 
coincidences between either proton recoil pulses from two 
neutrons or gamma-ray pulses from two neutron captures
occurring because of the high event rate. 
The good agreement between measured and simulated time delay distributions
gives evidence for the reliability of the
GEANT4 toolkit to model neutron interactions and capture
at MeV and sub-MeV energies. 

The expected time delay distribution is not purely exponential
since neutron capture occurred on several spatially separated
targets (hydrogen, Gd, copper, steel, lead etc.). For this reason, 
only the comparison between data and Monte Carlo is shown 
without any fit.

\subsection{Selection of muon events and capture gamma-ray pulses}

The data on muon-induced neutrons were collected from
August 2006 until April 2007. The veto detector was running in 
parallel with the ZEPLIN-II experiment. Time periods when the 
ZEPLIN-II detector (together with the veto system) was exposed 
to the calibration sources were excluded from the data analysis. 
The total live time of the experiment was 204.8 days. 

Muons were selected as follows: (i) trigger pulse area higher than
50~V$\times$ns corresponding to an energy threshold of
about 14~MeV (assuming proper reconstruction of the muon energy
deposition, i.e. no saturation of PMTs or DAQ); 
(ii) FWHM greater than 40~ns; (iii) time
during which the pulse amplitude exceeded 1.7~V, greater
than 10~ns. Only events with a trigger pulse area exceeding
70~V$\times$ns satisfy all selection criteria rising the energy
threshold to about 20~MeV (assuming no saturation). 
(In the pulse area -- to -- energy conversion above we used the more
reliable energy calibration using the 2.22~MeV gamma-rays from
neutron capture on hydrogen in the data run that will be described below.)
This allows rejection of
all background gamma-rays keeping more than 90\% of
muons \cite{muons,lindote07}. It also excludes a significant number of
events when low-energy
secondaries associated with muons are detected.
More details on muon flux simulations and muon
detection efficiency can be found in Refs.~\cite{muons,lindote07}.
During the experiment 10832 muons were detected translating to a rate of 
$52.9 \pm 0.5$ per day in agreement with
previous measurements \cite{muons}. Comparing the measured rate with
the Monte Carlo predictions gives 
the total muon flux at Boulby as 
($3.79 \pm 0.04$ (stat) $ \pm 0.11$ (syst))$\times 10^{-8}$~cm$^{-2}$~s$^{-1}$. 
The systematic error of the muon flux
is due to the uncertainty in the energy scale.
The above value is slightly smaller than reported in Ref.~\cite{muons}
mainly due to the more accurate three-dimensional Monte Carlo of muon
transport in the vicinity of and in the detector (with the account of all interactions).
In Ref.~\cite{muons} a one-dimensional simplified model was used.

The small reduction in the muon flux compared to previous measurements \cite{muons}
is equivalent to a small increase in the estimated depth
(column density) of the Boulby laboratory where the experiment was carried out. 
The vertical depth reconstructed from the present observations 
(assuming flat surface relief above the lab) is $2850\pm20$~m~w.~e.
(see Ref.~\cite{muons} for details on the depth reconstruction procedure).
Note that the present measurements of the muon flux have been carried out in
a new laboratory area at Boulby located about 200~m from the 'old' lab where
previous experiment \cite{muons} was performed. 
The muon flux reported here and in Ref.~\cite{muons} for the
Boulby Underground Laboratory is defined as the flux through a sphere
with unit cross-sectional area. This definition may be different from those used
for some other underground laboratories.

Secondary, delayed, pulses were selected using the following criteria:
(i) energy higher than 2~V$\times$ns or 550~keV;
(ii) presence of logic pulse;
(iii) time delay relative to the muon pulse from 20~ns to 190~ns.

\subsection{Results}

The measured and simulated energy spectra of delayed pulses are shown in 
Figure~\ref{fig-spn}. The energy scale for the measured spectrum
was chosen to achieve the best visual agreement with the position
of the simulated peak that corresponds to a combination of full absorption 
and Compton edge of 2.22~MeV
gamma-rays from neutron capture on hydrogen --
the main target for neutron capture in this experiment.
When the energy calibration from the $^{60}$Co run is used,
the peak appears at the energy of about $2.8\pm0.1$~MeV. 
For neutron data, this peak serves as an
independent and more accurate energy calibration. This is due to different 
positions of gamma-induced events within the detector in the 
calibration and data runs. In the
gamma calibration run, the source was located between the
veto detector and the main ZEPLIN-II target, close to the bottom
of the veto where the light collection was unfavourable. In the data run, neutrons
from muons were captured evenly in the detector volume.
Hence the peak in the spectrum at around 2~MeV provides an alternative
and, to a certain extent, more reliable energy calibration.
Superimposing this spectrum onto the simulated one, taking into account the
smearing of energy deposition due to finite energy resolution,
provides the pulse area -- to -- energy conversion factor. This conversion
is shifted by about 20\% relative to the $^{60}$Co calibration data
due to the different location of energy depositions as explained
above. Using the peak from neutron capture, the energy thresholds for 
neutron capture gammas was determined as $0.55\pm0.10$~MeV, and that
for muons as $20\pm5$~MeV. The uncertainty in the energy resolution and 
the statistical uncertainty in the peak position are responsible for the error
in the conversion factor. This will be converted later on to the systematic uncertainty
in the neutron yield. This error also gives the systematic uncertainty in the
muon flux.

The time delay distribution of pulses in the events relative to
trigger (muon) pulses is presented in Figure~\ref{fig-tdn}
together with the simulations. Both experimental data sets
(with roof section on and off, hereafter called 'roof-on' and
'roof-off') were combined together. Details of the simulations will be
described below.
Only the time window 40-190~$\mu$s after the muon trigger
is shown for experimental data. We neglected all events in
0-20~$\mu$s time window because of the large number of afterpulses.
In addition we did not consider events that occur at 20-40~$\mu$s 
after the trigger. This is because the neutron capture at that time may
happen with a non-negligible probability on Gd (about 18\% according to our
simulations as described below) which is difficult to simulate
accurately due to two reasons. Firstly, 
Gd loading and its distribution within wax is not known precisely. 
Secondly, gammas from neutron
capture on Gd are not described accurately enough in the GEANT4 model
framework. To avoid ambiguities in the data interpretation, we restricted
the time window to 40-190~$\mu$s after the trigger (11\% probability of capture on Gd
according to our simulations). Neutron capture on Gd was included in our simulations
but the reduced time window allowed us to minimise possible errors
associated with aforementioned uncertainties.
We do not expect a significant change in the neutron rate
even if captures on Gd were not accurately simulated.

The total number of secondary (delayed) pulses with energy deposition 
greater than 0.55 MeV in the time window of 40-190~$\mu$s after the muon
trigger was measured as 1037 whereas the number of muon triggers
was 10832 giving the rate of $0.096\pm0.003$ pulses/muon.

\begin{figure}[htb]
   \includegraphics[width=15.cm]{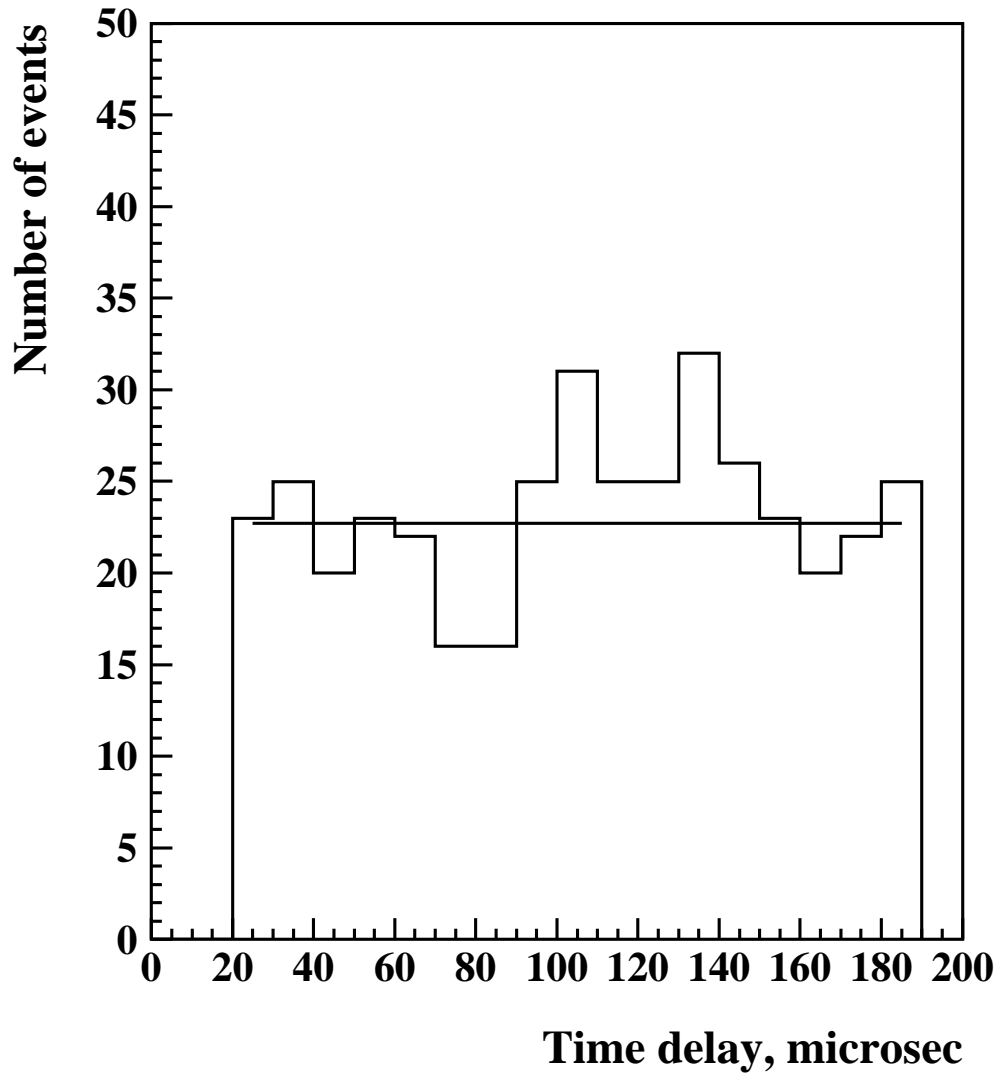}
    \caption{Time delay distribution of secondary pulses in gamma-induced events. 
    The data (histogram) were fitted
    to a constant background. Good agreement between data and flat fit 
    proves that the origin of these events is the random coincidences between
    background pulses.}
  \label{fig-tdb}
\end{figure}

A similar graph
for gamma events (small energy deposition, small FWHM and
no truncated amplitude) is shown in Figure~\ref{fig-tdb}. 
The distribution is flat proving
that the delayed pulses in gamma events are due to rare
random coincidences whereas the quasi-exponential 
shape of delayed pulses in muon events
is due to neutron captures. 

Random coincidences between
gamma background pulses should also be present in muon events
with the same rate as in gamma-induced events.
The distribution shown in Figure~\ref{fig-tdn} is in fact the sum
of an exponential (or several exponentials) due to neutron capture
and a flat background component due to random background
coincidences. Gamma-induced events allow us to determine
the flat background component or the rate of background
pulses in muon events. The number of background events in
40-190~$\mu$s time window after the gamma trigger 
(Figure~\ref{fig-tdb}) is 351 for 21461 gamma triggers giving
a rate of background pulses of $0.0164 \pm 0.0009$ per 
event. So the true rate of neutron pulses in muon events is the difference
between the total rate of secondary pulses in these events and the rate
of pulses due to random background coincidences (flat component).
Hence the neutron rate, defined
as the average number of detected neutrons per detected muon, 
is obtained as $0.079 \pm 0.003$ neutrons/muon
in 40-190~$\mu$s time window. The threshold for energy deposition from
neutron capture gammas was 0.55 MeV. The error is purely statistical. 
Systematic uncertainty is mainly due to the uncertainty in the pulse area -- to -- energy
conversion and will be added to the predicted neutron yield.

Figure~\ref{fig-mult} shows multiplicity distributions for
both types of events (muons -- solid histogram, and gammas --
dashed histogram), namely the number of events as a function
of the number of delayed pulses in an event. Both
distributions are normalised to the total number of events of a 
particular type. There is a bigger fraction of muon events with 
non-zero secondary (neutron) multiplicity, than the fraction of
gamma events (non-zero multiplicity is due to random coincidences). 
Large neutron multiplicity is expected in some muon events due to 
enhanced neutron production in lead, proving the neutron origin of
delayed pulses. Measured distribution has a maximum multiplicity of 16.
This number is smaller than the maximum multiplicity cut described in Section 2,
since in the final analysis the reduced time window of 40-190~$\mu$s
was used compared to data reduction procedure (15-190~$\mu$s).

For about 40\% of the exposure the detector was running without the roof of the castle
reducing the lead target mass for neutron production and the mass of 
Gd-loaded wax for neutron moderation and capture.
The two sets of runs (roof-on and roof-off) were analysed separately to
estimate the effect of the roof. The random background event rates
were determined separately from corresponding time distributions
of gamma events. Because of the absence of the roof section 
of the 'castle' (shielding)
the mean number of secondaries per gamma event 
with roof off was almost twice that number with roof on.
After subtraction of random background component
the neutron rates for two sets of runs were obtained as: 
$0.084 \pm 0.004$ neutrons/muon (roof-on) and $0.072 \pm 0.005$ neutrons/muon
(roof-off).

\section {Monte Carlo simulations}
\label{sim}

Measurement of the neutron rate using liquid scintillator at Boulby is hard to interpret
without the full Monte Carlo that had been developed using GEANT4 toolkit.
The presence of several potential targets for neutron production and neutron capture
requires a detailed model of the detector, its surroundings and physics processes.
Similar considerations are true for any other experiment with similar goals.
The main advantage of our measurements is in the use of a single detector with
well controlled energy threshold and systematics 
that makes its simulation a relatively easy task compared to larger modular 
detectors.

The modeling of the muon-induced neutrons has been carried out in two stages. 
In the first stage, muons were sampled using the muon generator MUSUN
\cite{vak03}
according to their energy spectrum and angular distribution at the 
Boulby Underground Laboratory.
These distributions were obtained by propagating muons from
the surface through the Boulby rock using the code MUSIC
\cite{music}. The calculated absolute muon flux was normalised to the
present muon flux measurements which, as mentioned above, 
agree well with earlier work \cite{muons}. 
The mean muon energy at Boulby was calculated as $\approx 260$~GeV.
Muons were sampled on the surfaces of a rectangular parallelepiped that
surrounded the main experimental hall of the underground laboratory
where the detector was placed. The distance from the 
surface of parallelepiped to the cavern walls was 7~m on four sides,
10~m above and 5~m below the detector. This ensured that high-energy
neutrons produced in rock far away from the laboratory hall and the
detector could still reach it and be
detected (for more discussion about neutron production by muons see
Ref. \cite{ha05}). Two million muons were generated in this way and their
parameters (energy, position, direction cosines and sign) were recorded
and passed to GEANT4 on the second stage.

Neutron production, transport and detection were simulated using GEANT4
\cite{geant4} version 8.2. The laboratory hall, lead and wax shielding, Gd, veto
and ZEPLIN-II detector were included in the simulations.
Detailed description of the simulations, various validation tests
and comparison with other simulation work will be presented in a separate
paper \cite{lindote07}. Here we present only the results important
for the interpretation of the experimental data.

The physics models and particle production thresholds ('cuts') were essentially 
those described in Ref.~\cite{ha05}, although other configurations were studied
as described in Ref.~\cite{lindote07}.
All hadronic and electromagnetic processes were taken into
account in the simulations.
The hadronic interactions were simulated using
the quark-gluon string
model above 6~GeV, an intra-nuclear binary cascade model at lower
energies and a pre-equilibrium de-excitation stage below 70~MeV. 
Neutron interactions below 20~MeV were treated using
high-precision data-driven model. The production thresholds (`cuts')
considered in these simulations were a few tens of keV for gammas and
$\sim$1~MeV for electrons and positrons in all materials. No thresholds were
applied to neutron tracking.

Muon parameters were read from the MUSUN output file and passed to
GEANT4 for further muon transport. Stochastic muon interactions and
continuous energy loss due to ionisation were simulated. All muons and
secondary particles
produced by the muons and in muon-induced cascades (showers) were 
transported and their energy depositions in the veto detector and interaction times
were stored in memory. Finally, for each muon event
we recorded the energy depositions for different time bins covering 
0--1~$\mu$s (10 bins), 1--200~$\mu$s (199 bins) and 
200--500~$\mu$s (30 bins). This allowed us to select events imposing
the same cuts as for real data sets and plot distributions in a similar
way as for the data run for direct comparison.

In total, about 120 million muons were sampled, each of the pre-recorded muons
was sampled about 60 times, but transported differently using different random 
numbers. This corresponded to a run time of about 
960 days.
The statistics are about 4.7 times that for real data.

\section {Discussion}
\label{discussion}

\begin{figure}[htb]
   \includegraphics[width=15.cm]{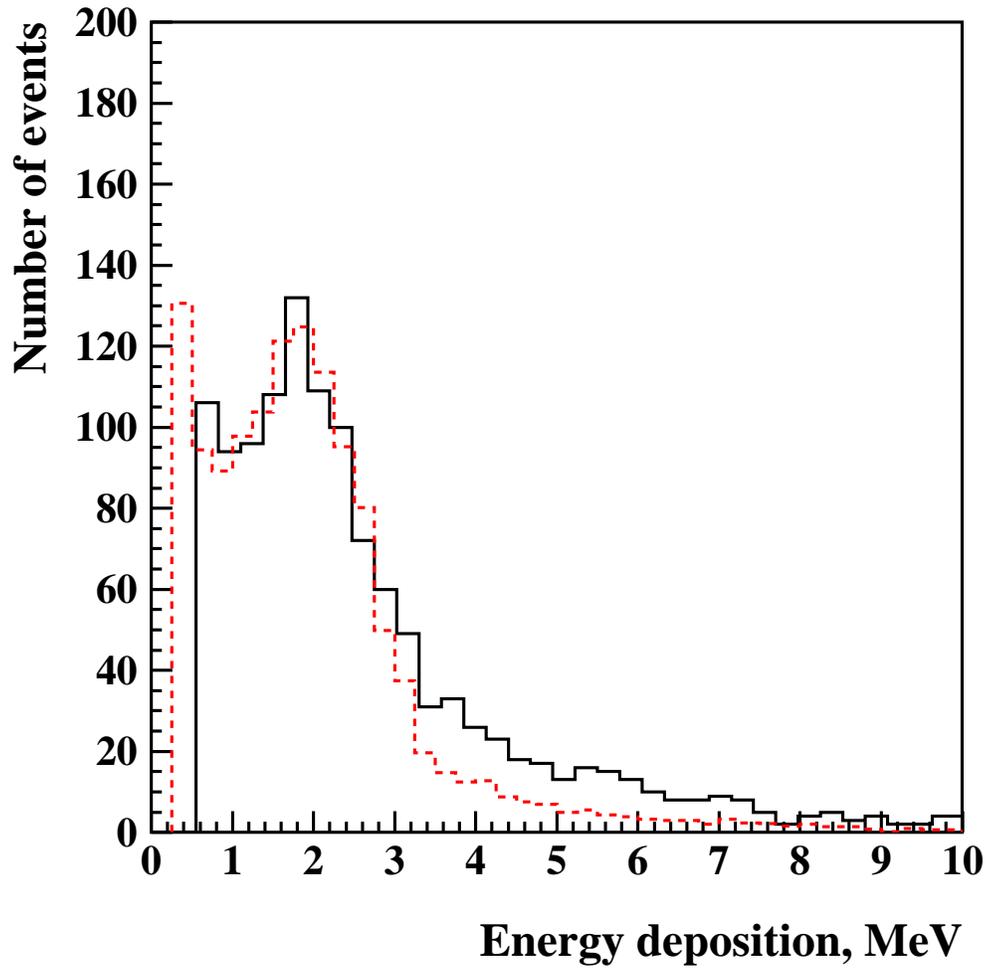}
    \caption{Energy spectrum of secondary pulses in muon events. 
    The data (solid histogram) are shown together with the simulated spectrum
    (dashed histogram) taking into account the energy resolution (Gaussian smearing).}
  \label{fig-spn}
\end{figure}

\begin{figure}[htb]
   \includegraphics[width=15.cm]{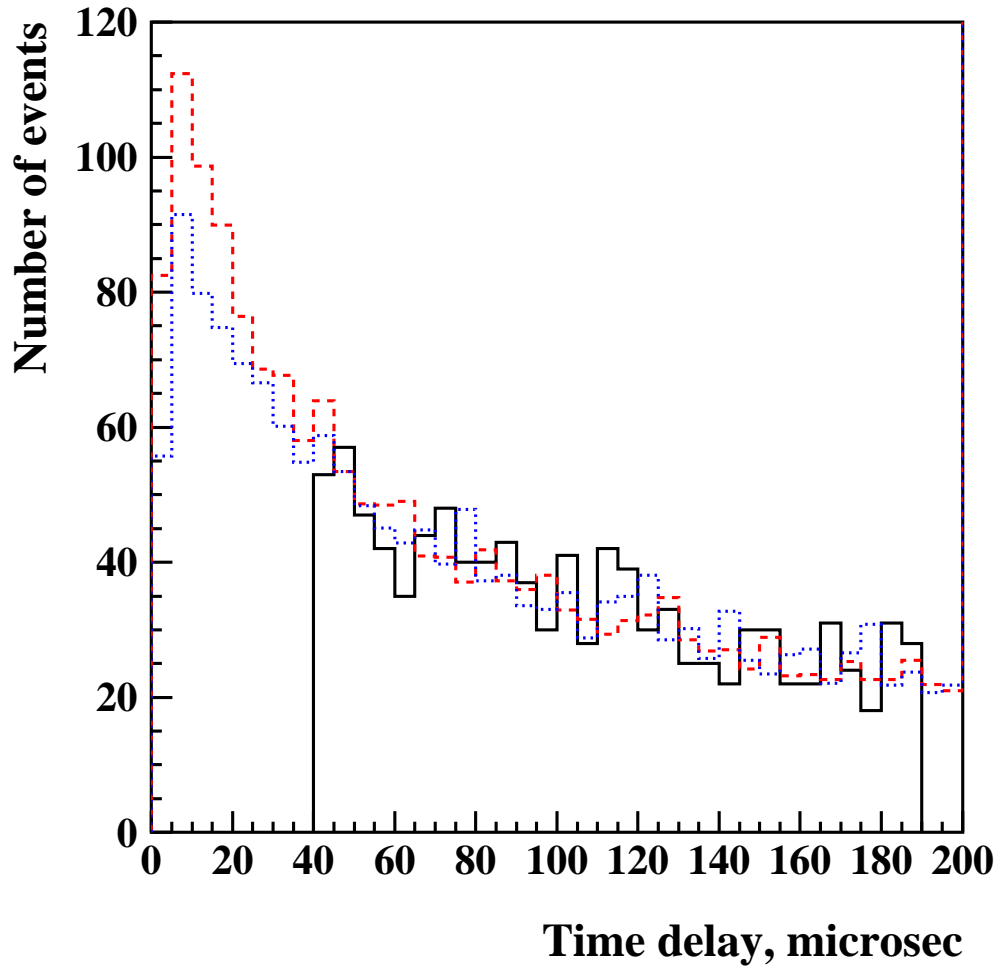}
    \caption{Time delay distribution of secondary (delayed) pulses in muon events. 
    The data (all runs combined -- solid histogram) are shown together with simulations
    with (dashed histogram) and without (dotted histogram) roof section.
    Simulations are normalised to the data using two free parameters: absolute
    normalisation and contribution from flat random background (see text for details).}
  \label{fig-tdn}
\end{figure}

\begin{figure}[htb]
   \includegraphics[width=15.cm]{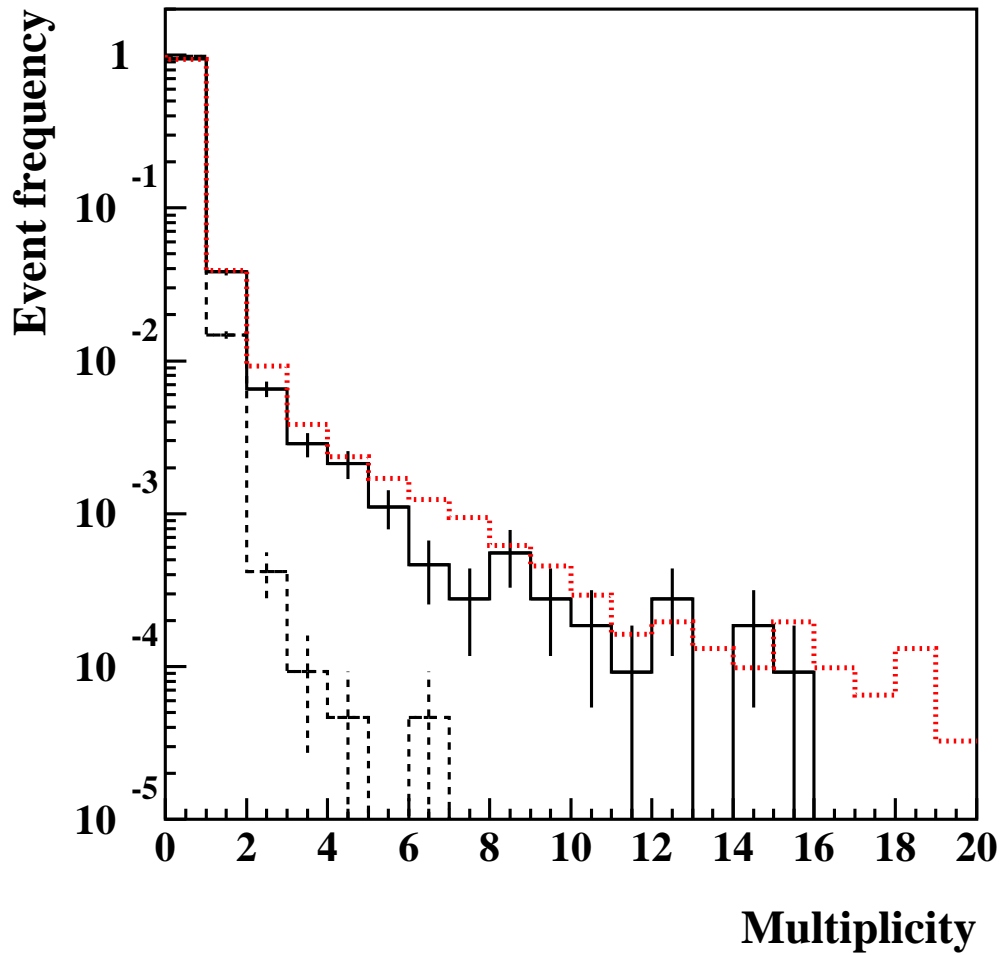}
    \caption{Measured multiplicity distributions of secondary pulses in muon (solid histogram)
    and gamma (dashed histogram) events. The histograms have been
    normalised to the total number of events of each type. First multiplicity
    bin corresponds to zero secondaries. Dotted histogram shows simulated
    distribution for runs with roof section on.}
  \label{fig-mult}
\end{figure}

Comparison between data and simulations is shown in Figures
\ref{fig-spn}, \ref{fig-tdn}, \ref{fig-mult}. Reasonable agreement between measured and
simulated spectra of delayed gammas, shown in Figure \ref{fig-spn}, allowed us
to establish the energy scale
and evaluate the threshold for muons and delayed gammas.
We estimated the 
uncertainty in the energy scale as about 20\% that leads to an uncertainty
in the energy threshold for delayed gammas of about 0.1~MeV. 
Changing the pulse area -- to -- energy conversion or energy resolution by this factor
destroys the agreement in the peak shape and position. 
This does not 
exceed the shift in the energy scale if the $^{60}$Co spectrum is used for
energy calibration, confirming the scale of the systematic uncertainty.
The uncertainty in the energy threshold for muons may be a little higher since
a saturation effect can suppress slightly the measured energy deposition 
from muons. We estimated the total systematic uncertainty in the energy scale
for muons as about 25\%. This is consistent with agreement between muon rates
reported here and in Ref.~\cite{muons} where the suppression of energy deposition
for muons occured at much higher energies. Large
suppression of muon energy deposition at threshold, leading to higher 
effective energy threshold, would manifest itself in a smaller muon rate measured
in the present experiment. Systematic uncertainties in energy scales for muons and
delayed gammas were converted into the uncertainty in the simulated neutron rates
and will be reported below.

Figure \ref{fig-tdn} shows the simulated time delay distributions compared to the
experimental data described above. Two sets of simulations (roof-on and roof-off)
are plotted separately to show good agreement in shape at 40-190~$\mu$s used
for data analysis. At small time delays a visible difference is due to the presence of
Gd in the roof shielding in the run with the roof section resulting in the enhanced
capture of neutrons on Gd at small times. 

The same energy thresholds for muon and delayed gamma energy depositions 
were applied to the simulated events as for data. 
The energy resolution was chosen to match the energy
spectrum of delayed gammas (see Figure~\ref{fig-spn}).
As for the neutron calibration run, tuning of simulated spectra was done using two 
free parameters: i) the total number of neutrons; ii) the flat component due to random
background coincidences (not simulated). The second parameter is bound by 
$\pm 3$ standard deviations from the measured background rate based on 
gamma-induced events (see Section \ref{results}). Figure~\ref{fig-tdn} shows
good agreement between measured and simulated shapes of the time delay
distribution. Note that the absolute normalisation was chosen to reach good
visual agreement and hence no conclusion about absolute neutron
rate can be drawn based on this figure.

The rate of simulated pulses from neutron capture gammas in 40-190~$\mu$s
can be directly compared to the measurements. This does not require any
assumption about the shape of the time delay distribution other than a general 
agreement between measured and simulated distributions (shown in 
Figure~\ref{fig-tdn}). Measured and
simulated neutron rates are compared in Table~\ref{tab1} for the two runs
separately and combined together. Only statistical errors are shown
in the table. Systematic uncertainty is mainly due to the 20\% uncertainty in the
energy scale as discussed above.
This was converted into the uncertainty of 0.009 neutrons/muon in the simulated
neutron rate in the 40-190~$\mu$s time window with a threshold of 0.55 MeV
for delayed pulses. Table~\ref{tab1} shows a factor of 1.8 difference
between measured and simulated neutron rates for both runs, GEANT4
predicting higher neutron rate than measured in the present experiment.

Figure~\ref{fig-mult} shows the simulated neutron multiplicity distribution 
(runs with roof-on) compared
to the data (all runs). A smaller rate of events is observed for almost all neutron
multiplicities. No more than 20 pulses on each waveform were recorded in the data 
run.
Since a restricted time window of 40-190~$\mu$s was used in
the analysis, the effective multiplicity cut was found to be 16 delayed pulses
(see Figure~\ref{fig-mult}). Taking the fraction of simulated events with
multiplicity larger than 16 we estimated the fraction of missed neutrons if the
maximum multiplicity cut of 16 was applied, as less than 2\%. This is smaller
than the statistical error of the measurements and does not affect the results.

A smaller observed neutron rate compared to simulations
can be due to two main factors: i) smaller neutron
yield than predicted in GEANT4 models; ii) enhanced absorption of neutrons or gammas
before neutrons or neutron capture gammas can reach the detector. The second argument
is highly unlikely because of the following reasons. Firstly, the measured time delay
distribution matches well the simulated one for the time window chosen. 
The shape of the time delay distribution depends on the neutron
capture cross-section and, hence, is affected by the geometry and neutron capture
models. Good agreement between data and simulations proves that the neutron capture
and secondary gammas are described reasonably well by the GEANT4 models.
Secondly, a similar agreement in shape was seen in the neutron calibration data proving
again a good model for neutron capture and gamma transport and detection.
Finally, neutron transport at MeV and sub-MeV energies was simulated with GEANT4
and MCNPX \cite{lemrani} and they were found to be in good agreement. Hence,
the most probable explanation of the observed deficit of muon-induced neutrons
lies in the GEANT4 model(s) involved in neutron production.

The ratio of roof-off / roof-on neutron rates is consistent with unity for both data (within
two standard deviations) and simulations.
Lead in the roof section contributes about 12\% to the total mass of lead.
Lead is responsible for about 91\% of neutrons in the
run with roof-on and about 90\% of neutrons in the run with roof-off
(results from our simulations).
The fact that neutron rates are similar in both sets of data despite
a 12\% difference in lead mass can be explained by the presence
of Gd-loaded wax in the roof section for runs with roof-on.
This enhances neutron moderation and capture above the detector
preventing neutrons (gammas) being detected in scintillator in runs
with roof-on. Thus, enhanced neutron production in lead in the roof
section was compensated by the efficient absorption of these neutrons
before they can reach the scintillator. Since lead contributes about 90\%
to the detected neutron rate, we can conclude that
GEANT4 probably overestimates the neutron yield in lead.
This result is somewhat surprising since GEANT4 was found to underproduce
neutrons relative to FLUKA Monte Carlo \cite{ha05}.

Alternative GEANT4 hadronic models result in similar neutron yields.
For the most widespread GEANT4 models the difference is within 20\%
\cite{lindote07,bauer}.
Recently developed CHIPS models \cite{kosov}
predict even higher flux
of neutrons and, hence, are probably inconsistent with the present
measurements. More detailed
studies will follow \cite{lindote07}.

The time window of 40-190~$\mu$s used in the data analysis contains about 42\%
of the total number of neutrons according to the simulations 
(41\% with roof section and 44\% without roof section), assuming a well
known geometry. Hence, the total neutron rate for an infinite time window can be 
evaluated as $0.188 \pm 0.005$ neutrons/muon for the simulated geometry. This figure is still 
geometry dependent. For practical use this figure has to be converted to the
neutron yield in a specific material, defined as the average number of neutrons
produced by a muon along 1~g/cm$^2$ of its path. The most reliable and quasi model
independent way of doing this is based on the following assumptions:
1) The geometry, neutron moderation and capture are described 
correctly in GEANT4; this is proven by the time delay distributions of data and  
calibration neutrons and by the agreement reached in Ref.~\cite{lemrani} for MeV and
sub-MeV neutron transport; 2) Most neutrons (about 90\%)
detected in our scintillator are produced in lead - this follows from simulations; 
3) The ratio
of measured-to-simulated neutron rates per detected muon
is thus the same as for raw neutron yields in 
lead. Using the fact that the measured neutron rate is 1.8 times smaller than
GEANT4 predictions, and the calculated neutron yield in the same model in 
lead is $2.37\times 10^{-3}$ neutrons/muon/(g/cm$^2$) for a muon energy of 260 GeV, we 
found the yield in lead, reconstructed from our measurements, to be 
$(1.31 \pm 0.06) \times 10^{-3}$ neutrons/muon/(g/cm$^2$), 
for a mean muon energy of about 260~GeV. This value
is 3.6 times smaller than the expected neutron yield from FLUKA simulations
\cite{ha05}.

Although our lead target and detector are of relatively small size, the fractional 
contribution of physical processes to the measured neutron rate does not differ
much from contributions of these processes 
to the total neutron yield (see, for instance, \cite{vak03,ha05}
for discussion), allowing us to convert our measured rate to the total neutron yield
as described above.
Our simulations show
that 2.2\% of detected neutrons are coming from the muon spallation
process, 63\% were initiated by photons or electrons and 34.8\% were produced 
by hadrons either via capture or hadron 
inelastic scattering. Although most neutrons detected
in our scintillator were produced in lead, cascades could be originated also in the
rock around the setup, thus increasing the fraction of detected neutrons from 
photons and hadrons.

The hypothetical contribution of a muon-induced neutron background for 
dark matter searches was checked using data from the ZEPLIN-II science run 
\cite{z2,z2-sd} with 225 kg$\times$days exposure. ZEPLIN-II
was operated in coincidence with the active veto system described here.
The data did not contain any neutron-induced nuclear recoil event in the target
(ZEPLIN-II) in coincidence with a large muon signal in the active veto
showing that for detectors with a few kilograms of target mass
this background is not a severe threat at a depth of about
3~km~w.~e.

\section {Conclusions}
\label{conclusions}

The first measurements of the muon-induced neutron flux at the Boulby
Underground Laboratory were presented. The experiment was carried out with an 0.73~tonne
liquid scintillator. The delayed coincidence method was used to detect a muon
(the first pulse in an event) and gamma-rays resulted from neutron capture on hydrogen
or other elements (secondary, delayed, pulses). The muon-induced neutron rate was
measured as $0.079 \pm 0.003$(stat.) neutrons/muon using
neutron-capture signals above 0.55~MeV in a time
window of 40-190~$\mu$s after the muon trigger. 
Accurate Monte Carlo simulations of the
neutron production, transport and detection in a precisely modeled laboratory and
experimental setup were carried out using the GEANT4 toolkit. The simulations
gave a rate of $0.143 \pm 0.002$(stat.)$\pm 0.009$(syst.) neutrons/muon for the
same selection criteria. The simulated result is 1.8 times higher than the measured value
and the difference largely exceeds all statistical and systematic uncertainties.
As the vast majority of neutrons detected in the current setup were produced in lead
we evaluated from our measurements the total neutron yield in lead as
$(1.31 \pm 0.06) \times 10^{-3}$ neutrons/muon/(g/cm$^2$)
for mean muon energy of about 260~GeV.

\section{Acknowledgments}
This work has been supported by the ILIAS integrating activity
(Contract No. RII3-CT-2004-506222) as part of the EU FP6 programme 
in Astroparticle Physics. We acknowledge the financial support from 
the Particle Physics and Astronomy Research Council (PPARC, currently Science
and Technology Facility Council -- STFC), the US Department of Energy (grant numbers
DE-FG03-91ER40662 and DE-FG03-95ER40917), the 
US National Science Foundation (grants
PHY-0139065 and PHY-0653459) and
from Funda\c{c}\~ao para a Ci\^encia e Tecnologia 
(project POCI/FP/81928/2007).
We are grateful to the Nuffield Foundation for the support of J. Blockley
through Undergraduate Research Bursary (grant URB/34132/2007).
We also wish to thank Cleveland Potash Ltd for assistance.

\pagebreak

\begin{table}[htb]
\caption{Measured and simulated rates of delayed pulses due to neutron capture
given in neutrons/muon:
2nd column (roof-on) -- for runs with roof section; 3rd column (roof-off) --
for runs without roof section; 4th column (ratio) -- ratio of rate without roof section
to that with section; 5th column (combined) -- the data from all runs are combined together.
Delayed pulses with energy deposition greater than 0.55 MeV
were counted in the 40-190~$\mu$s time window after the
muon trigger. Only statistical error is 
shown for simulation results. Systematic uncertainty of simulations is largely dominated
by the energy scale uncertainty (pulse area -- to -- energy conversion factor) and is equal to 0.009
for the two sets of runs and for the combined data. For the
ratio of rates in the two runs, the systematic uncertainty cancels out.}
\label{tab1}
\vspace{1cm}
\begin{center}
\begin{tabular}{|c|c|c|c|c|}\hline
& Roof-on & Roof-off & Ratio & Combined 
 \\ \hline
 Data & $0.084 \pm 0.004$ & $0.072 \pm 0.005$ &
$0.86 \pm 0.07$ & $0.079 \pm 0.003$ \\
\hline
 Simulations & $0.143 \pm 0.002$ & $0.143 \pm 0.003$ &
$1.00 \pm 0.03$ & $0.143 \pm 0.002$ \\
\hline
\end{tabular}
\end{center}
\end{table}

\pagebreak

\pagebreak

\pagebreak

\pagebreak

\pagebreak

\pagebreak

\pagebreak

\pagebreak

\pagebreak

\end {document}